\documentclass[useAMS,usenatbib]{mn2e}
\input epsf
\usepackage{graphicx}

\title[A precessing jet model for the PN K 3-35]{A precessing jet
  model for the PN K 3-35: simulated radio-continuum emission}
  \author[{Vel\'azquez et al.}]{Pablo
  F. Vel\'azquez$^{1}$\thanks{E-mail:pablo@nucleares.unam.mx}, Yolanda
  G\'omez$^{2}$\thanks{E-mail:y.gomez@astrosmo.unam.mx}, Alejandro
  Esquivel$^{1}$\thanks{E-mail: esquivel@nucleares.unam.mx}, and Alejandro C. Raga$^{1}$\thanks{E-mail:raga@nucleares.unam.mx}\\ $^{1}$Instituto de
  Ciencias Nucleares, Universidad Nacional Aut\'onoma de M\'exico,
  Ciudad Universitaria,\\ Apartado Postal 70-543, CP 04510, M\'exico
  D.F., M\'exico.\\ $^{2}$Centro de Radioastronom\'\i a y Astrof\'\i
  sica, Universidad Nacional Aut\'onoma de M\'exico,\\  Apdo. Postal 3-72
  (Xangari), CP 58089, Morelia, Michoac\'an,
  M\'exico.}
\begin{document}

\date{Accepted . Received ; in original form }

\pagerange{\pageref{firstpage}--\pageref{lastpage}} \pubyear{2002}

\maketitle

\label{firstpage}

\begin{abstract}

  The bipolar morphology of the planetary nebula (PN) K 3-35 observed in
  radio-continuum images was modelled
  with 3D hydrodynamic simulations with the adaptive grid
  code {\sc yguaz\'u-a}. We find that the observed morphology of this PN
  can be reproduced considering a precessing jet evolving in a dense
  AGB circumstellar medium, given by a mass loss rate $\dot{M}_{csm}=5\times
  10^{-5}~\mathrm{M_{\odot} yr^{-1}}$ and a terminal velocity
  $v_\mathrm{w}=10~\mathrm{km s^{-1}}$.
  Synthetic thermal radio-continuum maps were generated from numerical
  results for several frequencies. Comparing the maps and the total
  fluxes obtained from the simulations with the observational results, we
  find that a model of precessing dense jets, where each jet injects 
  material into the surrounding CSM at a rate $\dot{M}_j=2.8\times 10^{-4}~\mathrm{M_{\odot}\ yr^{-1}}$ (equivalent to a density of 
  $8 \times 10^{4}~\mathrm{cm}^{-3}$), a velocity of 
  $1\,500~\mathrm{km~s^{-1}}$, a precession period of 100~yr, and a
  semi-aperture precession angle of 20\degr\ agrees well with the 
  observations.

\end{abstract}


\section{Introduction}
\label{sec:intro}
The evolution between the end of the asymptotic giant branch (AGB) and
the planetary nebula (PN) phases has for a long time
been a poorly understood link
in the late stages of intermediate-mass stars (1 -- 8 M$_\odot$).
It is in this transition phase where the fast stellar wind of the emerging PN
interacts with the slow wind from its precursor AGB star (Kwok, Purton, \&
Fitzgerald 1978), shaping the final morphology of the PN.

Other
ingredients that can contribute to the final shape of a PN are the presence of
interacting binary stars (e.g.  Morris 1987; Balick \& Frank 2002;
Soker \& Bisker 2006) or magnetic fields (e.g. Garc\'\i a-Segura \& L\'opez
2000; Blackman et al. 2001; Garc\'\i a-Segura, L\'opez \& Franco 2005).
Thus, the study of objects in this transition phase can give us
important clues about the physical mechanisms responsible for the different
morphologies observed in PNe.

K 3-35 is a very young PN with a characteristic S-shaped emission morphology
that suggests the presence of precessing bipolar jets (Aaquist \& Kwok 1989;
Aaquist 1993; Miranda et al. 2000; 2001).  The detection of OH and H$_2$O
maser emission as well as the presence of CO and HCO$^+$ emission suggest that
K~3-35 departed from the proto-PN phase only a few decades ago (Miranda et al.
2001; Tafoya et al. 2007).

Miranda et al. (2001) estimate a dynamical age of $\leq$ 15 years
for the ionised inner core, which expands at $\sim$25 km~s$^{-1}$.
For the jets, assuming a modest jet velocity of $\sim$100 km~s$^{-1}$, an age
of $\sim$800 years is obtained. Therefore, the jet formation in K~3-35
occurred during the proto-PN phase.
In this paper we show that the radio morphology of K 3-35 can be
explained by a precessing jet model.

\section{The precessing jet model}

\begin{figure} 
\begin{center}
\includegraphics[width=\columnwidth]{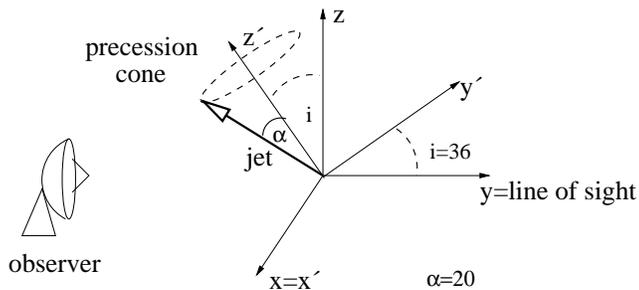}
\caption{Scheme of the initial conditions for the simulation. The $xyz-$~frame
corresponds to the computational domain. The physical system is given by the
$x'y'z'-$~frame, which is obtained by rotating the $yz$-plane around
$x$-axis by 36\degr with respect to the plane of the sky ($xz$-plane).}     
\label{fcroquis}
\end{center}
\end{figure}

The PN K 3-35 belongs to the point-symmetric PN group because it shows a
bipolar structure, with an `S' shaped morphology in radio-continuum maps at
8.3~GHz \citep{lfm01,lfm98,aaquist93,gomez03}.

This kind of morphology is more often observed in galactic and
extragalactic jets. Several theoretical studies have employed precessing jets
for modelling these objects and explaining the `S' shape observed at different
frequencies.

For instance, in the galactic case, 
\citet{masciadri02} carried out 3D numerical simulations of the HH~34
jets, and compared jet models with and without precession.
In the field of extragalactic jets, \citet{rodriguez06} made a
3D hydrodynamical simulation of a precessing jet moving into an
intragalactic cluster medium, and compared their results with the
Abell 1795 galaxy cluster.
The similarity between these jets and point-symmetric PNe has recently
motivated models of this type of PNe as jets, considering precession,
density and/or variable ejection velocity, etc., in order to explain
their kinematics, emission, and morphology. An excellent review of
these works is given by \citet{soker06}.

As an example, we can mention the
observational study of the proto-planetary nebula (PPN) 
Hen 3-1475 by \citet{riera03}. In H$\alpha$, [NII], and [OIII] images, this PPN
exhibits several knots forming an `S' shape. Also \citet{riera03} made
radial velocity and proper motion studies, showing that the knotty structure
of Hen 3-1475 can be explained by means of a jet with a variable ejection
velocity and precession. Based on these hypotheses, \citet{vel04} have
computed 2D (axisymmetric) and 3D hydrodynamical simulations of
such a jet, showing that this model can reproduce both the
observed morphology and the kinematical characteristics of the outflows of Hen
3-1475. 

In the present paper, we model the PN K 3-35 with 3D numerical
simulations of a precessing jet. Due to the fact that the bipolar
structure of K 3-35
looks almost continuous, we do not find necessary 
to consider also a time variability for the jet velocity and/or density.

In PNe, the surrounding circumstellar medium (CSM) is 
strongly modified by the central star, which injects material and increases
the density of the CSM.
Based on the work of \citet{mellema95}, this perturbed CSM
can be described by:

\begin{equation}
\rho(R)=\rho_0 \biggl[1-\delta \bigg({{1-\exp(-2 \beta \sin^2\theta)}
\over{1-\exp(-2\beta)}}\biggr)\biggr] (R_0/R)^{2}\, ,
\label{agb}
\end{equation}  

\noindent where $R=\sqrt{x^2+y^2+z^2}$ is the distance from the jet source
on a  3D Cartesian frame. The angle $\theta$  is measured with
respect to the equator, which we have chosen to coincide
with the $xy$-plane, yielding $\theta=\tan^{-1}(\sqrt{x^2+y^2}/z)$.
Eq. \ref{agb} describes the stratification of the surrounding CSM 
produced by a star in AGB phase. 
The parameter $\delta$ in Eq. \ref{agb} is related to the ratio of 
the density at the equator ($xy$-plane) and at the pole ($z$-axis), and
 $\beta$ determines how the density varies from the equator to the 
pole \citep{mellema95}.
The value of $\rho_0$ is calculated from the mass loss rate as~:
${\rho_0=\dot{M}/(4\pi\,R_0^2\,\mathrm{v_{env}})}$, where
$\mathrm{v_{env}}$ is the expansion velocity of the AGB remnant.

\begin{figure} 
\begin{center}
\includegraphics[width=\columnwidth]{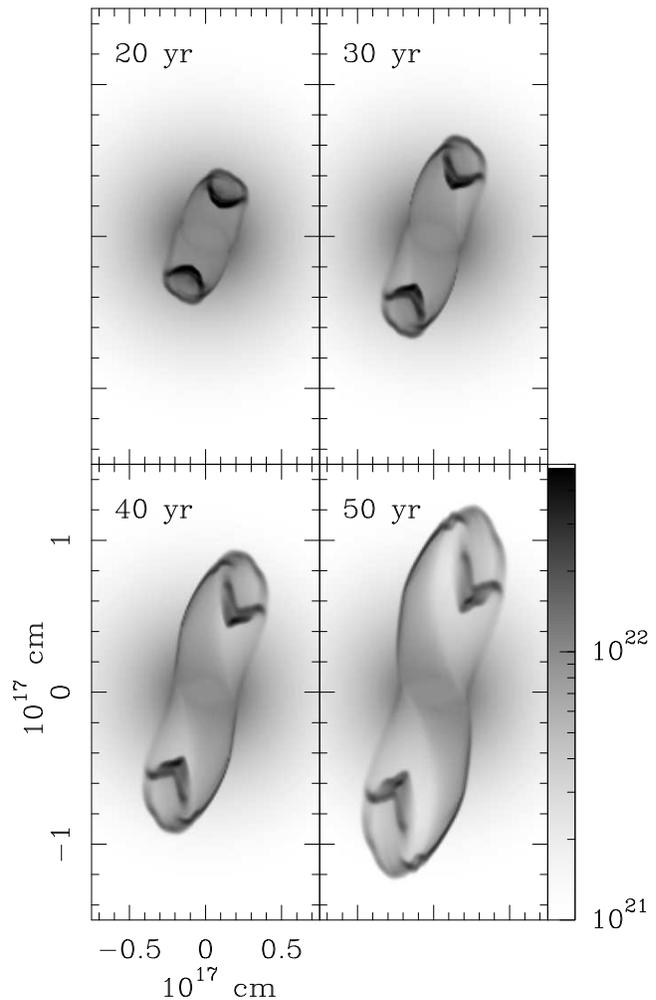}
\caption{Temporal evolution of the column density at integration times
  $t=20,~30,~40,~50$~yrs. These maps were obtained
  integrating the density along the $x$-axis. 
At t=40~yr, the jet achieves an extension of 
$1.8\times 10^{17}$~cm, which is equivalent to 2.4\arcsec , i.~e., the angular
  extension of bipolar structure of PN K 3-35 (for an assumed distance of
5 kpc). Both axes are labelled in units of $10^{17}$~cm, and
the bar on the right represents the logarithmic gray-scale, given in units of
  cm$^{-2}$.  }     
\label{f1}
\end{center}
\end{figure}

\section{The numerical simulations}
\subsection{Initial numerical simulation setup}
\label{sec:conditions}

The numerical simulations were carried out employing the 3D adaptive
grid code {\sc yguaz\'u-a} \citep{raga00}. Together with the gas-dynamic
equations, several
 rate equations for atomic/ionic species (H {\sc i-ii}, 
He {\sc i-iii}, C {\sc ii-iv}, and O {\sc i-iv}) are integrated in order to
calculate a non-equilibrium ionisation cooling function for the gas 
(for temperatures above $\sim 10^5$~K the cooling
rate is approximated by an equilibrium ionisation parameterised function).
The gas-dynamic equations are integrated with the ``flux-vector splitting'' 
algorithm of \citet{vanleer1982}.
Details of the reaction and cooling rates included in the code are described in
 the appendix of \citet{raga2002} and \citet{raga2007}. 

The simulations were performed in a five level, binary adaptive grid,
with a maximum resolution of $5.9\times 10^{14}$~cm (along the
three axes). The computational domain had an extent of 
$1.5\times 10^{17}$~cm in the $x$- and $y$-directions, and
of $3\times 10^{17}$~cm in the $z$-direction.

Both the cone precession axis and the CSM density distribution
were rotated around the $x$-axis, by 36\degr with respect to the plane of the
sky ($yz$-plane, see Fig.\ref{fcroquis}).
This inclination angle was found by  \citet{uscanga07}, who modelled
the {\sc h$_2$o} maser emission of the accretion disk around the
K 3-35 central star. The configuration is shown in Fig. \ref{fcroquis}.

The jet is placed at the centre of the computational domain with an
initial radius $r_\mathrm{j}= 1.5\times 10^{16}$~cm (equivalent to 30
pixels at the maximum grid resolution) and a length
$l_\mathrm{j}=6\times 10^{15}$~cm. Initially (integration time
$t=0$~yr), the jet axis lying on the $x'z'$-plane and starts to move
(counterclockwise) into a precession cone, with a semi-opening angle
$\alpha=20^{\circ}$. At $t=0$~yr, the jet velocity components are
$v'_j=(464.,0., 1426.)\ \mathrm{km\ s^{-1}}$, in the $x'y'z'-$~frame
(see fig. \ref{fcroquis}), or $v_j=(464., -764., 1204.)\ \mathrm{km\
s^{-1}}$, in the $xyz-$~frame (the computational domain frame).  After
several tests, a precession period of $\tau_p=100$~yr was chosen. 
Two runs were computed considering that the jet material is injected
as singly ionised or neutral gas. K~3-35 has a bright centre when it is
observed at radiofrequencies, because the central source strongly
ionises the CSM gas. This process is not taken into account in our
simulations.  Since we continuously impose the jet inflow conditions
within a cylindrical volume (see \S 2), when we calculate
synthetic radio-continuum maps we obtain a strong central emission
component (coming from this cylinder) for the initially ionised jet
simulation, giving a good morphological agreement with the
observations. However, this central component in the simulated radio
emission is exclusively a result of the chosen jet inflow conditions,
and for this reason only synthetic maps obtained for the run with the
``neutral jet'' will be employed for calculating total fluxes.

A good fit with the observations is obtained when the velocity and mass loss
rate of the jet were set to $1,500\ {\rm{km\ s}}^{-1}$ and
$\dot{M}_j=2.8\times10^{-4} \mathrm{M_{\odot}\ yr^{-1}}$, respectively. These
parameters let an equivalent density $\rho_j$ of $8 \times
10^{4}~\mathrm{cm}^{-3}$. The neutral CSM is modelled by Eq.\ref{agb}, with a
mass lose rate $\dot{M}=5\times 10^{-5}\, \mathrm{M}_{\sun}\,\mathrm{yr}^{-1}$
, $v_\mathrm{w}=10^{6}\,{\rm{cm\ s}}^{-1}$, and the parameters $\delta$ and
$\beta$, were set to 0.1 and 3, respectively. These parameters give an almost
isotropic density distribution with a contrast of 20\% between the density at
the equator and the poles. From this relationship, densities between
$7.5\times10^{-19}\mathrm{g\ cm^{-3}}$ and $1.7\times 10^{-20}\mathrm{g\ 
  cm^{-3}}$ are obtained for distances from the source of $1.5\times
10^{16}$~cm and $10^{17}$~cm along the $z'-$ direction, respectively.  These
high densities are supported by the recent detection of HCO$^+$ toward K~3-35,
that reveals the presence of high density molecular gas \citep{tafoya07}.

Taking into account the CSM density distribution, and the jet density and 
velocity, it is possible to estimate the speed of the leading `working surface'
(i.~e. the head) of the jet~:

\begin{equation}
v_\mathbf{ws}=v_{\mathrm{j}} \frac{\sqrt{\rho_\mathrm{j}/\rho_\mathrm{csm}(R,\theta)}}
                          {\sqrt{\rho_\mathrm{j}/\rho_\mathrm{csm}(R,\theta)}+1},
\label{vws}
\end{equation}

\noindent where $v_\mathrm{j}$ is the jet injection velocity, and
$\rho_\mathrm{j}$ and $\rho_\mathrm{csm}$ are the jet and CSM densities, 
respectively. It is straightforward to show from Eqs. (\ref{vws}) and
(\ref{agb})
that for large distances $R$ from the jet source, $v_\mathbf{ws}$ increases.
For instance, considering the values for $v_\mathrm{j}$, $\rho_\mathrm{j}$, and
$\rho_\mathrm{csm}$, speeds of $4.4\times 10^2 \rm{km\ s^{-1}}$ and
$1.1\times 10^3 \rm{km\ s^{-1}}$, are found at $R=5\times 10^{16}$~cm
and $R=10^{17}$~cm , respectively.

\subsection{Calculation of the thermal radio-continuum emission}
\label{genxray}

\begin{figure}
\includegraphics[width=\columnwidth]{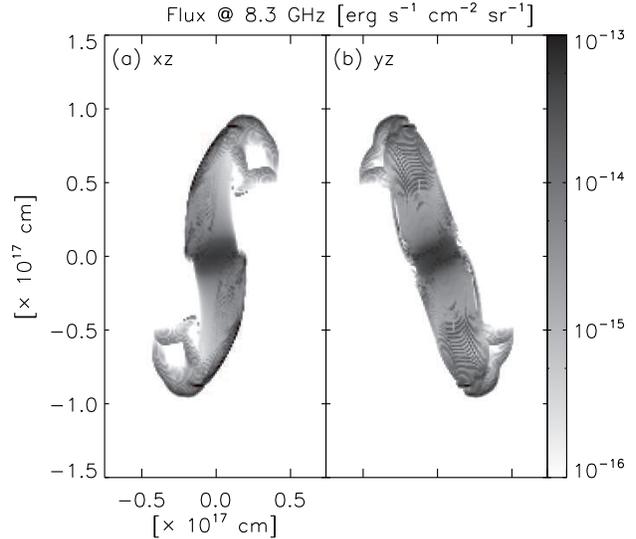}
  \caption{Simulated thermal radio-continuum maps obtained for a frequency of
    8.3 GHz. The left panel is the emission map generated by 
integrating the emission coefficient along the $y$-axis, the right panel
is the result of integrating along the $x$-axis.  
The bar on the right 
shows the logarithm gray-scale of the radio thermal emission in units of erg
 s$^{-1}$  cm$^{-2}$ sr$^{-1}$. Both spatial axis are in units of
$10^{17}$~cm. Strong emission is observed in the centre due to the fact 
that this is the region where a singly ionised jet inflow condition is 
continuously imposed.} 
\label{f2}
\end{figure}

To generate synthetic thermal radio-continuum maps, the
radiative transfer equation was integrated along different lines of
sight.

First, the optical depth $\tau_\nu(l)$ was calculated by integrating
the absorption coefficient $\kappa_{\nu}$ along lines of sight~:
$\tau_{\nu}(l)=\int_0^l \kappa_{\nu} dl$  (with $l$ being the
spatial coordinate along the line of  
sight). The absorption coefficient $\kappa_{\nu}$ is given by
\citep{rl79,osterbrock89}:

\begin{equation}
\kappa_{\nu}=0.018\ T^{-3/2}\ 
n_\mathrm{e}\ n_\mathrm{i}\ \nu^{-2}\ g_\mathrm{ff},
\label{taunu}
\end{equation}
 
\noindent where $g_\mathbf{ff}=0.5513\,(ln(T^{3/2} \nu^{-1})+17.7)$ is the
Gaunt factor.
We obtain directly from our simulations $T$, $n_\mathrm{e}$, and
$n_\mathrm{i}$ (the spatial distributions of
temperature, electronic, and ionic number
densities, respectively), which are used in Eq. (\ref{taunu}) to
calculate $\kappa_{\nu}$  (in units of $\mathrm{cm}^{-1}$).

Then, the intensity as a function of frequency can be obtained as~:

\begin{equation}
I_{\nu}=\int^l_0 j_{\nu}(n_e,n_i,T)\ \mathrm{e}^{-\tau_{\nu}(l)} dl, 
\label{inu}
\end{equation}

\noindent where  $j_{\nu}$ is the emission coefficient (which is
a function of $T$, $n_\mathrm{e}$,  and $n_\mathrm{i}$).
The intensity $I_{\nu}$ has units of $\rm{erg\ s^{-1}\ 
sr^{-1}\ cm^{-2}\ Hz^{-1}}$.
Numerically, the emission coefficient can be
calculated using the following equation \citep{rl79,osterbrock89}:

\begin{equation}
j_{\nu}=5.4113\times 10^{-39} n_\mathrm{e}\ n_\mathrm{i}\ T^{-1/2}\
\mathrm{e}^{-h\nu /k T} g_\mathrm{ff},
\label{jnu}
\end{equation}

\noindent where $h$ and $k$ are the Planck's and Boltzmann's constants,
respectively. The coefficient $j_{\nu}$ has units of $\rm{erg\
  s^{-1}\ sr^{-1}\ cm^{-3}\ Hz^{-1}}$.

Synthetic maps at different frequencies were
generated combining the results from the 3D simulation
with Eqs.~(\ref{taunu}), (\ref{inu}), and (\ref{jnu}).
The angularly integrated fluxes (in $\mathrm{mJy}$)
corresponding to different frequencies were 
then determined from the simulated maps, assuming a distance to the 
source of 5~kpc
\citep{zhang95}.

\section{Results}

\begin{table}
\begin{center}
  \begin{tabular}{rccr}
            \hline
            \noalign{\smallskip}
     $\nu$ (GHz)  &  S$^{sim}_{40}$ (mJy)&  S$^{sim}_{50}$ (mJy)& S$_{obs}$(mJy)\\
             \hline
           \noalign{\smallskip}
1.5  &  21.7  &  13.4 & 11.8 to 18.4\\
2.4  &  20.6  &  15.5 & ---\\
5.0  &  20.0  &  18.6 & 22.0$\pm$2.2\\
8.3  &  17.8  &  19.9 & ---\\ 
10.0  &  17.4 &  20.0 & ---\\ 
15.0  &  16.4 &  19.8 & 24.0$\pm$2.4\\ 
22.0  &  15.5 &  19.1 & 22.0$\pm$2.2\\
           \hline
            \noalign{\smallskip}
  \end{tabular}
\caption{Total flux densities (in units of mJy) at different frequencies,
  obtained from the simulated images at $t=40$~yr (second column), and at
  $t=50$~yr (third column). In this calculation, a distance of 5 kpc has been
  assumed to PN K 3-35. These flux densities have been obtained from the
  simulation where the jet material is imposed as neutral gas, in order to
  avoid an unrealistic strong emission from the centre.  For a direct
  comparison with observations, the fourth column gives the ``halo'' observed
  fluxes, taken from \citet{aaquist93}.}
\end{center}

\label{tab1}
\end{table}

With the physical conditions described in \S 3.1, a 3D numerical simulation
was carried out. Figure \ref{f1} shows the temporal evolution of
the column density obtained for integration times lying in the range
[20,50]~yrs. The stratification of the surrounding CSM is clearly observed
in all maps. 

Filamentary structures are observed at the head jets, due to precession
movement. At different times, jet material is injected at different angles.
These jet material sweep up the CSM which have a density stratification as
function of the angle (see Eq. \ref{agb}), generating the bright filaments
observed in the column density maps.

At a $t=40$~yr integration time, the jet reaches a total
length of $1.8\times 10^{17}$~cm. This size is equivalent to 2.4\arcsec
(i.~e., the 
angular extension of K 3-35 on the plane of the sky), at a distance of
5 kpc.
\citet{lfm01} have made estimates of the dynamical ages of both the
inner ionised core and the extended emission (jets) of the PN K 3-35.
Based on an expansion rate of $\sim 25~\rm{km~s^{-1}}$ taken from the
He II lines, they obtained a dynamical age of $\le 15~\rm{yr}$ for the core, 
while assuming a jet velocity of $\sim 100~\rm{km~s^{-1}}$, a dynamical age
of $\sim 800~\rm{yr}$ is deduced for the jets.
Our simulations required a much larger jet velocity to explain the
radio emission of the lobes in the PN K 3-35, thus yielding a short 
dynamical age for the jets, which is comparable with that of the core.
This is interesting because it may suggests that the process that originated
the ionisation of the jets and the core occurred almost simultaneously.

Synthetic maps were generated (as described in section 3.2) 
for the following frequencies: 1.5, 2.4, 5, 8.3, 10, 15, and 22 GHz.
Very similar morphologies are obtained for all of these frequencies.
Figure \ref{f2} shows the maps obtained at 8.3 GHz.
The left panel of Figure \ref{f2} displays the simulated radio-continuum
emission when the $j_{8.3\mathrm{GHz}}$ (Eq. \ref{jnu}) is integrated
along the $y$-axis, giving the emission on the $xz$-plane.  
The right panel of Figure \ref{f2} shows the
8.3 GHz emission on the $yz$-plane (i.~e., the integration of
$j_{8.3\mathrm{GHz}}$ 
was carried out along the $x$-axis). The map on the right was shown
to illustrate how
the morphology strongly depends on the projection chosen. In the 
$xz$-projection, the radio
emission has an `S' shape (which is the morphology of K 3-35), while
in the $yz$-projection, we have a `Z' shape. These maps are displayed in
units of $\mathrm{erg\ cm^{-2}\ s^{-1}\ sr^{-1}}$.

Comparing the left panel of Figure \ref{f2} with the column density map at
$t=40$~yr of Figure \ref{f1}, we see that the radio-continuum
emission traces very well the precessing jet morphology.

It is also possible to compare simulated and observed total fluxes.
However, since we are not simulating the radio emission from the central
region, we can only compare the radio emission from the lobes or jets, which
were called the `halo' in \citet{aaquist93}.
From each simulated radio map we measured the corresponding total 
flux from these structures. These fluxes are listed in Table
\ref{tab1}, where one can see that the total fluxes in the simulations
(at all frequencies) are of the order of the observed ones.
At an integration time of 40~yr, the simulated flux gives an optically thin
behaviour for the spectrum. On the other hand, for integration times of 50~yr,
the simulated flux has a maximum at a frequency between 5 and 8.3 GHz, 
in good agreement with the Aaquist (1993) results. 
At frequencies higher than 8.3 GHz, the spectral index is of $-0.09$,
which corresponds to optically-thin thermal free-free emission.

A direct comparison between both observed and simulated 8.3 GHz maps
is shown in Figure \ref{fcomp}. The simulated map was convolved with
a synthesised beam of $0.2\arcsec \times 0.2\arcsec$ 
(similar to the beam in the observations), and scaled to yield units
of mJy beam$^{-1}$. The same scale for both images have been employed. 
Also, the synthetic map was tilted in 48\degr (with respect to $z-$~axis, see
Fig. 1)  in order to do the position angle (P.A) of the jet direction (in the
$xz-$~plane) equal to 65\degr, the observed P.A.\citep{lfm01}.

\section{Summary}

Several 3D hydrodynamical simulations were carried out employing the adaptive
grid code {\sc yguaz\'u-a}, in order to  model both the morphology and
the thermal radio 
emission of the planetary nebula  K 3-35.

After analysing our results, we found that the bipolar structure of this PN 
can be described as the
result of the interaction of a dense jet (with an initial number density
of $8\sim 10^4$~cm$^{-3}$ or $\dot{M}_j=2.8\times 10^{-4} M_{\odot}
yr^{-1}$) moving into a dense environment, previously swept up by the AGB wind of
the central star. 

The `S' morphology shown by  K~3-35 in 8.3 GHz radio-continuum images
\citep{lfm98,lfm01,gomez03} can be
reproduced if the modelled jet precesses with a period of 100~yrs on a cone
with a half-opening angle of 20\degr.

For an integration time of 40~yrs, the simulated jet has a total length of
$1.8\times 10^{17}$~cm, which is equivalent to 2.4\arcsec 
\citep{lfm01,gomez03}considering an estimated distance of 5 kpc to K 3-35 
and also the orientation of this object \citep{uscanga07}. This time is 
almost 20 times smaller than the one given by \citet{lfm01} for the jet, 
where a slower velocity was assumed. However, it is only $2.7$ times larger
than the dynamical age of the inner core \citep[also in][]{lfm01},
suggesting that the two events are more related than previously
thought, and further supporting the idea that K 3-35 is a young
object.

\begin{figure}
\includegraphics[width=\columnwidth]{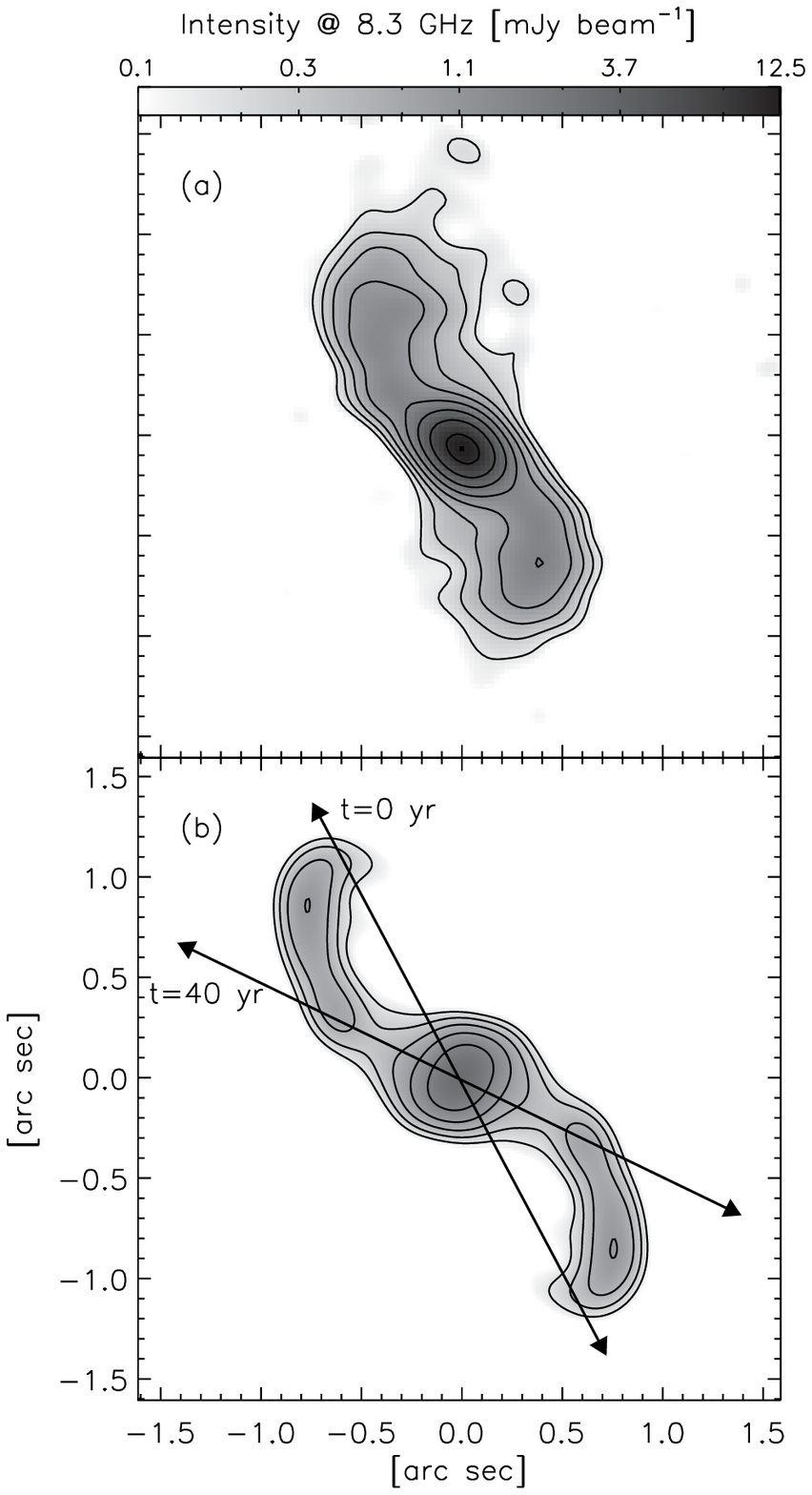}
  \caption{Comparison between observational and numerical results. The
    top panel shows the 8.3~GHz radio continuum image of K 3-35
    \citep{lfm01,gomez03}.  The bottom panel shows the tilted
    synthetic radio emission map at the same frequency, obtained for
    an integration time of 40~yr.  An inclination of 36\degr between
    the precession axis and the plane of the sky was considered. The
    simulated map was smoothed with the observed beam of
    $0.2\arcsec\times 0.2\arcsec$ and it is shown in the same scale
    that the observed one.  The contours corresponds to the levels
    0.18, 0.31, 0.56, 1.0, 1.78, 3.16, 5.62, and 10.00 mJy
    beam$^{-1}$. The arrows indicate the jet direction (projected on
    the plane of the sky) for integrations times of 0 and 40~yr. The
    last one is coincident with the P.A. measured by \citet{lfm01}.}
 \label{fcomp}
\end{figure}

Synthetic radio-continuum maps were generated from our numerical
results.  These maps show that the predicted morphologies and fluxes
are in reasonable agreement with the observations. At an integration
time of 40~yrs, the obtained spectral index is the one of optically
thin emission. For an integration time of 50~yrs, the observed change
of the spectral index with frequency (Aaquist 1993) is also reproduced
by our simulation.  A direct comparison between the observational and
numerical results is given in Fig. \ref{fcomp}, where we show the
observed and synthetic (for an integration time of 40~yrs) 8.3 GHz
radio maps.

We must note that the values for the velocity and mass loss rate
employed in the simulation for the jet and the CSM seem to be rather
high. It is difficult for AGB and post-AGB starts to launch jets at
velocities of the order of $1\,500\,\mathrm{km~s}^{-1}$ (although
Riera et al. 2003 have reported this kind of velocity for the outflows
of the PPN 3-1475).  Besides, the mass injection is also a bit
extreme, in 40 years both jets have injected $0.02~\mathrm{M}_{\odot}$
into the surrounding CSM. This scenario might be explained in terms of
a binary system, if the jet is produced by a companion accreting
material (at a rate about ten times higher than $\dot{M}_j$) from a
massive star (the AGB progenitor that produced the density
distribution of the circumstellar material). The primary in the last
40 years has lost $0.2~\mathrm{M}_{\odot}$. This means that it started
with an envelope containing this mass, which appears to correspond to
an AGB star rather than a post-AGB star (even though at the present
time the star has already evolved to the planetary nebula stage).

Furthermore, the CSM is very dense. With the parameters employed, in a
radius of $10^{17}$~cm, a mass of 1 M$_{\odot}$ is contained, implying
a massive PN. There is observational evidence that favours a
very dense surrounding CSM. However, the total mass derived
from HCO$^+$ observations (Tafoya et al. 2007) is quite low
($\sim 0.017~$M$_\odot$), so that the molecular emission from this molecule
would be confined to a small, possibly shock excited volume,
in order to be consistent with our much higher mass CSM. These values imply
that this scenario would be plausible if it is a short lived
event. Clearly, a better determination of these parameters could be
done with proper motion studies and better distance determinations.

Notwithstanding the extreme values for the  employed parameters, it is
important to note that a simple model of a precessing dense jet moving 
into an also dense CSM,
successfully reproduces the observed morphology of the PN K 3-35, obtaining
a qualitatively and quantitatively good agreement between the model
predictions and the observations, although this scenario would be a short lived
event.

\section*{Acknowledgements}

The authors acknowledge support from CONACyT grant
46628-F, and DGAPA-UNAM grants IN108207 and IN100407. 
The work of ACR, AE, and PFV was supported by the ``Macroproyecto
de Tecnolog\'\i as para la Universidad de la Informaci\'on y la
Computaci\'on'' (Secretar\'\i a de Desarrollo Institucional de la UNAM,
Programa Transdisciplinario en Investigaci\'on y Desarrollo
para Facultades y Escuelas, Unidad de Apoyo a la Investigaci\'on en
Facultades y Escuelas). Authors sincerely acknowledge Noan Soker (the referee) 
for his very useful comments, which allow us to improve the previous version of this manuscript.
We also would like to thank the computational team of ICN:
Antonio Ram\'\i rez and Enrique Palacios, for maintaining and supporting our Linux servers, and  
Mart\'\i n Cruz  for the assistance provided.
Finally, PFV acknowledges the hospitality, during his visits, of the CRyA (campus Morelia-UNAM) staff.

\bsp
\label{lastpage}
\end{document}